\documentclass[%
 aip,
 amsmath,amssymb,
 reprint,%
]{revtex4-1}

\usepackage{graphicx}
\usepackage{dcolumn}
\usepackage{bm}

\usepackage[utf8]{inputenc}
\usepackage[T1]{fontenc}
\usepackage{mathptmx}
\usepackage{etoolbox}

\makeatletter
\def\@email#1#2{%
 \endgroup
 \patchcmd{\titleblock@produce}
  {\frontmatter@RRAPformat}
  {\frontmatter@RRAPformat{\produce@RRAP{*#1\href{mailto:#2}{#2}}}\frontmatter@RRAPformat}
  {}{}
}%
\makeatother

\newcommand{\beginsupplement}{%
        \setcounter{table}{0}
        \renewcommand{\thetable}{S\arabic{table}}%
        \setcounter{figure}{0}
        \renewcommand{\thefigure}{S\arabic{figure}}%
     }

\begin{document}

\preprint{AIP/123-QED}

\title[]{Stress-induced omega ($\omega$) phase transition in Nb thin films for superconducting qubits}
\author{Jaeyel Lee}
\author{Zuhawn Sung}
\author{Akshay A. Murthy}
\author{Anna Grassellino}
\author{Alex Romanenko}
\affiliation{Fermi National Accelerator Laboratory (FNAL), Batavia, IL 60510, USA}

\email{jlee406@fnal.gov, amurthy@fnal.gov}

\date{\today}

\begin{abstract}
We report the observation of omega  ($\omega$) phase formation in Nb thin films deposited by high-power impulse magnetron sputtering (HiPIMS) for superconducting qubits using transmission electron microscopy (TEM). We hypothesize that this phase transformation to the $\omega$-phase with hexagonal structure from bcc phase as well as the formation of {111}<112>  mechanical twins is induced by internal stress in the Nb thin films. In terms of lateral dimensions, the size of the $\omega$-phase of Nb range from 10 to 100 nm, which is comparable to the coherence length of Nb ($\approx$ 40 nm). In terms of overall volume fraction, $\approx$ 1\% of the Nb grains exhibit this $\omega$-phase. We also find that the $\omega$-phase in Nb is not observed in large grain Nb samples, suggesting that the phase transition can be suppressed through reducing the grain boundary density, which may serve as a source of strain and dislocations in this system. The current finding may indicate that the Nb thin film is prone to the $\omega$-phase transition due to the internal stress in the Nb thin film. We conclude by discussing effects of the $\omega$-phase on the superconducting properties of Nb thin films and discussing pathways to mitigate their formation. 
\end{abstract}

\maketitle

Nb thin films have been extensively used for resonators and electrodes of superconducting qubits \cite{mcrae2020materials,romanenko2017understanding,murthy2022tof}. Scaling up this technology to deliver truly transformational quantum computing solutions to the larger community requires improving metrics such as the coherence times, which represent the period over which information remains in a state of quantum superposition. Specifically,  it has been observed that materials defects, such as interfaces, surfaces, and impurities play critical roles in inducing this unwanted quantum decoherence\cite{de2021materials,place2021new,murray2021material}. For instance, it has been shown that amorphous regions at various interfaces in the device act as parasitic two-level systems (TLS) and cause decoherence \cite{muller2019towards,pappas2011two,murthy2022potential}, including metal-air \cite{altoe2022localization,premkumar2021microscopic,romanenko2017understanding}, substrate-air \cite{altoe2022localization,earnest2018substrate}, and metal-substrate \cite{megrant2012planar,wenner2011surface}. Similarly, there have been reports on impurities and precipitate formation serving as possible sources of decoherence via   pair-breaking \cite{lee2021discovery,murthy2022tof}. Identifying the atomistic origins of microwave loss in these devices and mitigating them has rapidly developed into an active research area.

To date, investigations centered on structural changes in Nb thin film and their effects on superconducting properties have been limited. This topic requires further study, however, as Nb exhibits a low yield strength(30-80 MPa) and  critical resolved (20 MPa) shear stress, making it easily deformable under stress \cite{bieler2010physical}. Further, it has been shown that when various bcc metals such as Ta, Ti, and Zr experience stress, it leads to twinning and a subsequent phase transformation from bcc to omega ($\omega$) phase with hexagonal crystal structures \cite{xing2008mechanical,hsiung2000shock,sikka1982omega}. Whereas, there has been a recent report on the formation of $\omega$-phase transition in single crystal Nb when subjected to compressive stress \cite{li2021shear}, no such reports exist for Nb in the thin film geometry. Further, as Nb is soft with low yield strength as described, it is possible that internal stress in Nb thin film is sufficient to induce this metastable phase and merits further investigation \cite{chakraborty2011thickness,clemens1990metastable}.  

Here, we report the observation of {111}<112> twinning-induced $\omega$ phase formation of Nb thin film of superconducting qubits. We find that this twinning and $\omega$-phase formation in Nb occurs in as-grown Nb thin film on Si thus indicating internal stress in the Nb thin film is sufficient to induce this phase transformation. Previous XRD analysis indicates that the Nb thin films are under internal stress\cite{murakami1985strain} and, which may possibly explain the twinning and $\omega$-phase formation in the as-grown Nb thin films on Si substrate. $\omega$ phase transformation is not seen in the bulk Nb, indicating that it is likely that the grain boundary plays a role in the formation of twinning and $\omega$-phase transition. Possible effects of $\omega$-phase transition in the superconducting properties of Nb thin films are discussed.

\begin{figure}

   {\includegraphics[width = 2in]{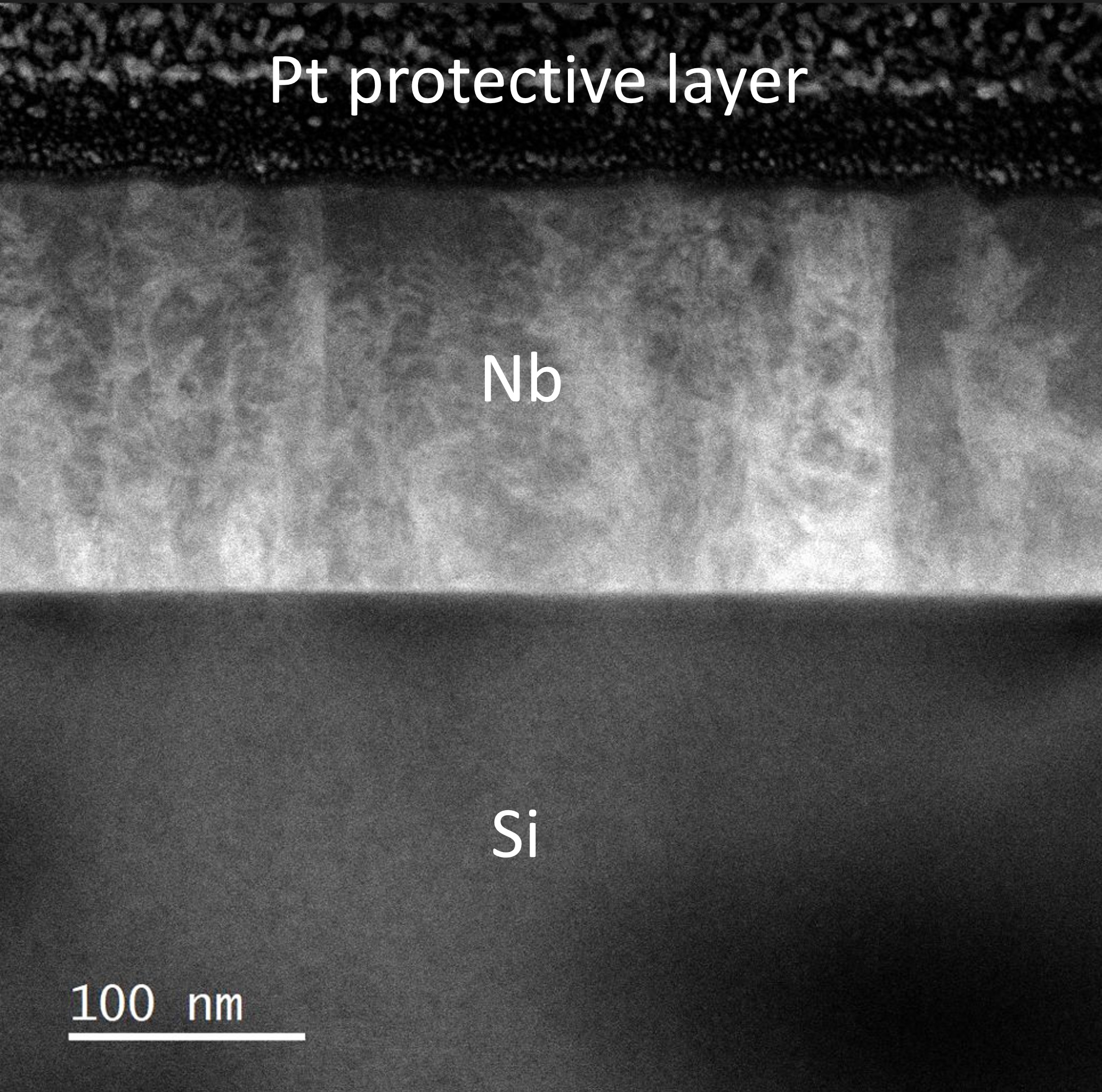}}
   \caption{ADF-STEM image of 160 nm thick Nb thin film on Si (100) substrate deposited by HiPIMS. It shows typical columnar grain structures of Nb thin films on Si substrates for films deposited with this method.}
\end{figure}

\begin{figure*}[!t]
   \centering
   {\includegraphics[width = 6 in]{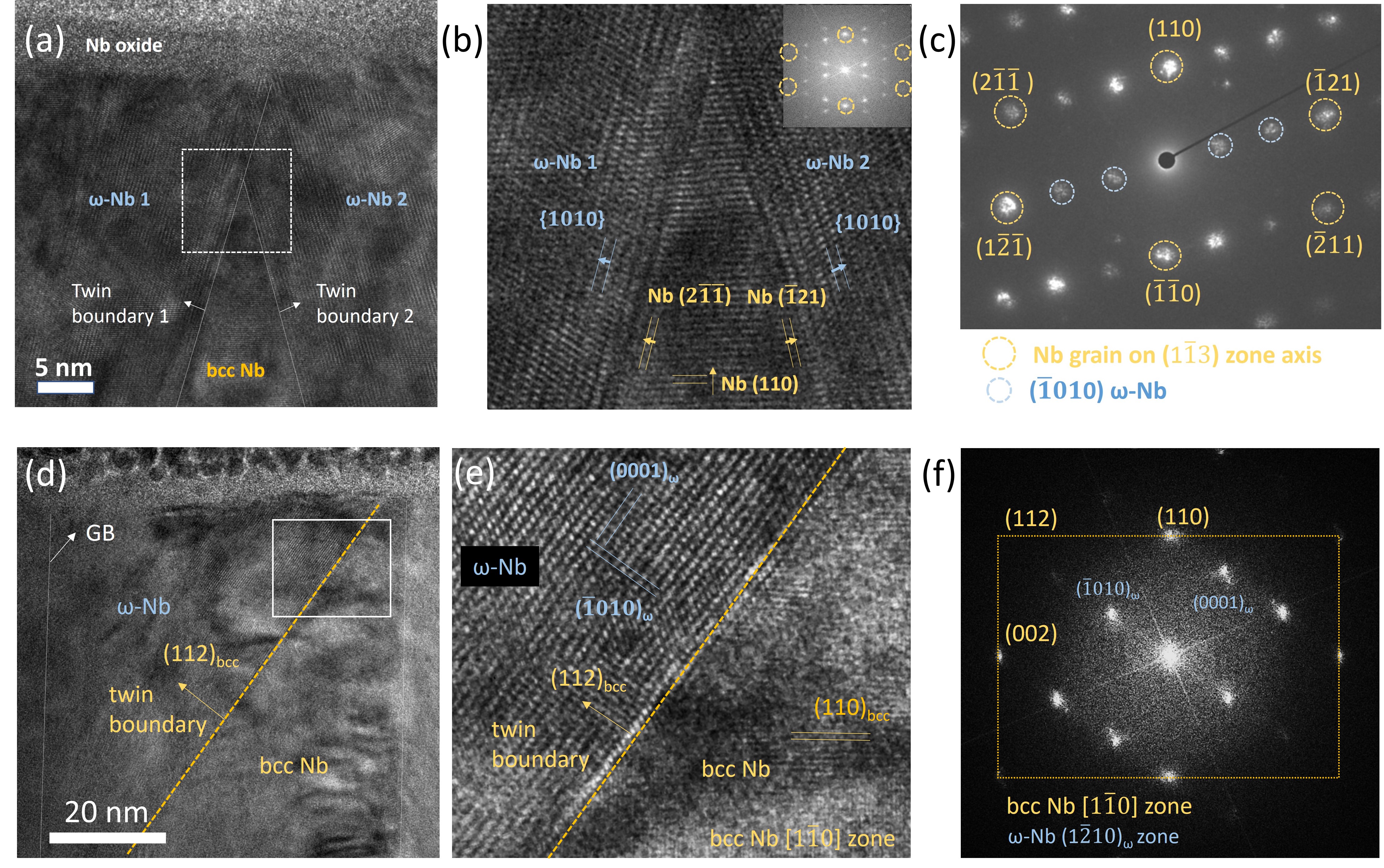}}
   \caption{(a) HR-TEM image of Nb grain with two {112}<111> twin boundaries with fully transformed $\omega$-Nb phases on bcc Nb [11$\bar{3}$] zone axis and (b) the details of the two twin boundaries denoted by the square in Fig. 1(a) are shown. FFT of the region with $\omega$-Nb displays the reflections from bcc Nb [1$\bar{1}$3] zone axis indicated by brown circles and additional reflections from (1010){$_{\omega}$} of the two $\omega$-Nb phases are also seen. (c) Nanobeam electron diffraction were taken on the ${\omega}$-Nb phase 2 on bcc Nb [11$\bar{3}$] zone axis and it also show additional ($\bar{1}$010){$_{\omega}$} reflections from the $\omega$-Nb phase.(d) HR-TEM image of Nb thin films display the $\omega$-Nb phase in Nb [1$\bar{1}$0] zone axis. (e) The magnified HR-TEM image from the area in Fig. 1(d), which is denoted by white rectangular, is displayed. (f) FFT of the HR-TEM image of $\omega$-Nb phase in Nb demonstrate that the appearance of reflection from $\omega$-Nb phase. 3-dimensional orientation relationships (ORs) of $\omega$-Nb phase in Nb are observed,  [1$\bar{1}$0]$_{bcc}$//[1$\bar{2}$10]{$_{\omega}$} and (112)$_{bcc}$//($\bar{1}$010){$_{\omega}$}, which is the typical orientation relationships for $\omega$-phases in bcc metals.}
\end{figure*}

170 nm thick Nb thin films are deposited on Si (100) substrates using high power impulse sputtering (HiPIMS) at Rigetti computing. TEM samples of Nb thin films were prepared by focused ion beam (FIB) with 30 kV Ga ion beams and damaged layers on the surface of the Nb thin foil are removed using 5 kV and 2 kV Ga ion beams. High-resolution TEM analyses were performed using JEOL ARM 200 CF microscope with 200 kV electron beam. The microscope was equipped with a Cold FEG source and probe aberration corrector. ADF images were acquired using a convergence semi-angle of 21 mrad and collection angles of 68-260 mrad. TEM images are recorded and analyzed using Gatan Micrograph Suite software. Superconducting characterizations using typical 4-point electromagnetic characterization is performed in physical property measurement system (PPMS) to estimate the critical temperature and residual resistivity ratio (RRR).

Annular dark-field (ADF)-STEM image of Nb thin film on Si substrate in Fig. 1 displays 170 nm thick Nb thin film on Si (100) substrate. It shows that Nb grains have columnar structures with ~50 nm of grain diameter. HR-TEM images of the Nb thin film are provided in Fig. 2 and exhibit the {112}<111> twin, which we observe frequently throughout the Nb thin film and may be induced by the internal stresses in the Nb thin film \cite{murakami1985strain}. In conjunction with these {112}<111> twins on the surface, we observe the presence of $\omega$-Nb in the surrounding regions. HR-TEM images in Fig. 2(a-c) illustrate the presence of $\omega$-Nb near the surface of the Nb film along with the formation of {112}<111> twins on bcc Nb [11$\bar{3}$] zone axis. Nanobeam diffraction taken from $\omega$-Nb is provided in Fig. 2(c) and displays reflections associated with the (1010){$_{\omega}$} planes parallel to (112)$_{bcc}$. We note that bcc Nb and $\omega$-Nb form a sharp interface along the {112}<111> twin boundary.

In Fig. 2(d-f), HR-TEM images of another $\omega$-Nb phase on Nb [110] zone axis are provided. In the HR-TEM image, we also observe that $\omega$-Nb and bcc Nb form a sharp interface along the {112}<111> twin boundary. Associated FFT patterns of the HR-TEM images across the $\omega$-Nb and bcc Nb phase demonstrate the appearance of reflections from $\omega$-Nb along with bcc Nb matrix. This satisfies the typical 3-dimensional orientation relationships of $\omega$-phase in bcc matrix: [1$\bar{1}$0]$_{bcc}$//[1$\bar{2}$10]{$_{\omega}$} and (112)$_{bcc}$//($\bar{1}$010){$_{\omega}$}\cite{sikka1982omega}. HR-TEM image in Fig. S1 shows that $\omega$-phase transition of Nb also occurs near the Nb/Si interface. In Fig. S1, (112) twin boundary is seen when Nb is tilted to the [113] zone axis and bcc Nb transforms to $\omega$-phase across the twin boundary.

\begin{figure*}
    {\includegraphics[width = 6 in]{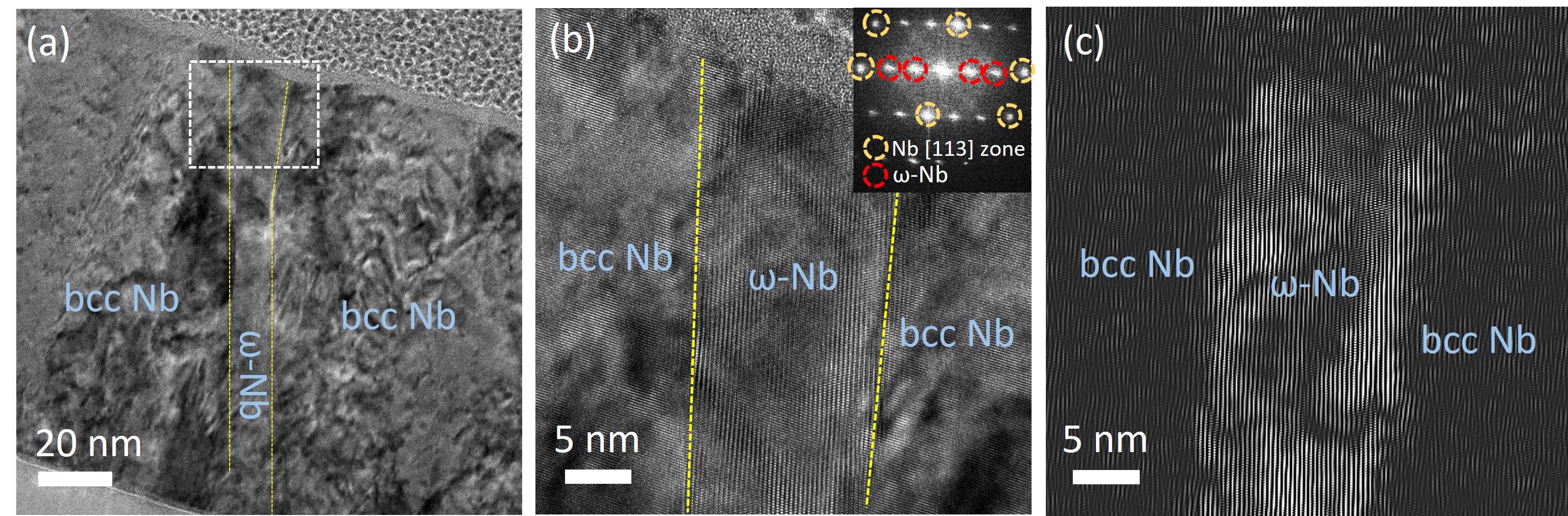}}
   \caption{(a) A low magnification overview of the omega ($\omega$) phase in Nb thin film. (b) A high magnification HR-TEM image of a selected area from the surface to substrate direction in Fig. 3(a). An associated FFT pattern taken from the $\omega$-phase is provided as an inset and reflections from bcc Nb [113] zone axis are denoted by yellow dotted circles. Additional reflections are seen inside the reflections from the bcc Nb on [113] zone axis. (c) Inverse FFT image is taken from the $\omega$-Nb reflections in the FFT of Fig. 3(b) and the morphology of the $\omega$-Nb is displayed.}
\end{figure*}

In TEM image provided in Fig. 3(a), a lower magnification overview of the morphology of $\omega$-phase along the depth-direction of the Nb thin film is illustrated. Here, we observe that the $\omega$-phase extends along the entire film thickness (170 nm). A magnified HR-TEM image of the $\omega$-phase near the surface on Nb [113] zone axis is displayed in Fig. 3(b), showing the details of the interface between bcc Nb and $\omega$-phase Nb. The corresponding FFT pattern is provided as an inset and shows the appearance of additional peaks associated with the $\omega$-phase. To more clearly illustrate the spatial location of the $\omega$-phase, an inverse FFT image using signal associated only from these additional peaks is provided in Fig. 3(c).  

As discussed previously, while this phase has not received significant attention in the case of Nb metals, it has been widely observed in bcc and hcp metals such as Zr, Ti, and Ta alloys \cite{sass1972structure}. In this case, the size of the $\omega$-Nb phases ranges from 10 to 100 nm and volume fraction of the $\omega$-Nb phase in the bcc Nb thin film is estimated to be $\approx$1\%.
As the $\omega$-phase transition has been previously linked to stress\cite{hsiung2000shock, li2021shear}, the existence of $\omega$-phase in Nb thin film indicates that such stresses are present in the Nb thin films. Lattice mismatch at Nb/Si interface likely serve as one source of such stresses in the Nb thin films. Additionally, a high level of O, C, H impurities near the surface of Nb may cause additional stresses and cause twinning and $\omega$-phase transition\cite{murthy2022tof}. Impurity levels of hydrogen and oxygen in the top surface of Nb is roughly anticipated to be 1-3 at.\% \cite{murthy2022tof} and it may cause 0.2-0.5\% of strain on the surface of Nb thin film. As the Young’s modulus of Nb is 104 GPa\cite{antoine2011physical}, the additional stresses caused by impurities in Nb is 200-500 MPa, which is well above the yield strength and critical resolved shear stress of Nb. These two sources of stresses may help explain why the $\omega$-phase is frequently observed near the surface of Nb as well as the metal/substrate interface.   
   
The superconducting properties of the $\omega$-Nb phase require further investigation. But as electron-phonon coupling may change in $\omega$-Nb compared to bcc Nb \cite{savrasov1996electron,zarea2022effects}, it is anticipated that the superconducting properties such as superconducting gap ($\Delta$) and critical temperature (T$_{c}$) may also differ in $\omega$-Nb compared to bcc Nb. For instance, there has been a report that the presence of substantial volume fractions of $\omega$-phase (>5 vol\% according to TEM micrographs) leads to a changes of T$_{c}$ from 10.2 K to 6.8 K in Zr-15at.\%Nb alloys\cite{narasimhan1972superconducting}. Also, Ti-4.5at.\%Mo and Ti-10at.\%Mo alloys show that the formation of $\omega$-phase in Ti-Mo alloys degrades T$_{c}$\cite{ho1969influence}. For bcc Nb, the electron-phonon coupling constant is 0.83 and the critical temperature is 9.3 K,  assuming that Debye temperature ($\theta$$_{D}$) and effective electron mobility ($\mu$) are 275 K and 0.13, respectively\cite{mcmillan1968transition}. For Nb under non-equilibrium state, it can possibly form fcc structure \cite{sasaki1988fcc,oh2022multi} and in this case, the electron-phonon coupling constant has been estimated to be 0.53, which leads to a decrease in the critical temperature of Nb down to 2.3 K \cite{sasaki1988fcc}. Further, recent studies imply that a slight deviation from bcc structure of Nb leads to a decrease in the density of states near the Fermi level which, in turn, induces a decrease in a critical temperature of Nb \cite{privatecommunicationl}. These reports suggest that the $\omega$-phase of Nb may also possibly lead to a local degradation in superconducting properties such as low T$_{c}$ compared to bcc Nb phase, which can lead to Cooper pair breaking and quantum decoherence in superconducting qubit systems.   

\begin{figure}
    {\includegraphics[width = 3 in]{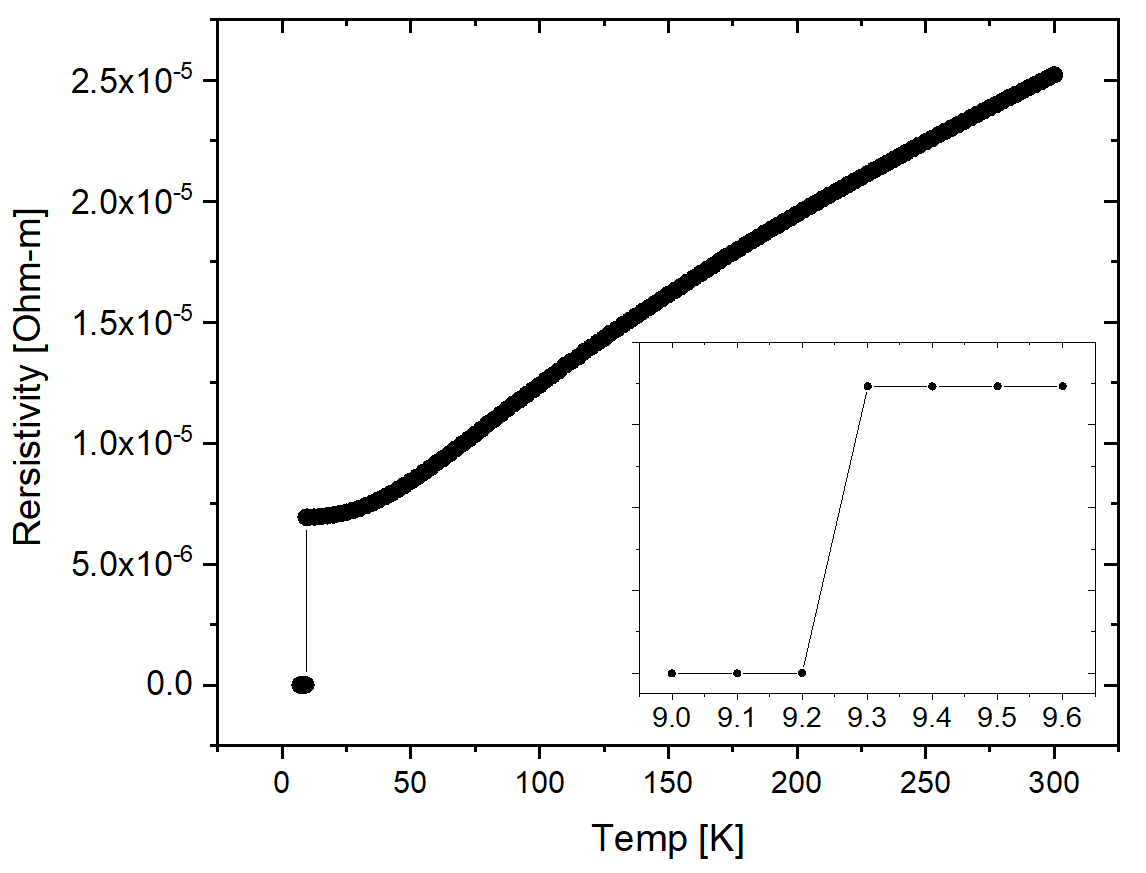}}
   \caption{The plot of resistivity vs temperature for the Nb thin film on Si (100) substrate, acquired by conventional 4-points measurement. }
\end{figure}

Due to the limited volume fraction of $\omega$-phase ($\approx$1 vol.\%,), physical property measurement system (PPMS) measurements are not well-suited to identify the relationship between this phase and T$_c$. Namely, we observe that this Nb thin film exhibits a critical temperature of Nb thin film is 9.3 K and displays a sharp transition from normal state to the superconducting state, where $\Delta$Tc (T$_{90\%}$- T$_{10\%}$) is 0.08 K from the resistivity measurement in Fig. 4. The value of residual resistivity ratio (RRR) is estimated to be 3.64 similar to typical polycrystalline Nb thin films\cite{bose2006upper}. We also note that preliminary magnetometry measurement indicates that the coherence length of the Nb thin film, $\xi$(0), is possibly shorter than the typical value of a high purity Nb, $\xi(0)$ $\approx$ 40-50nm and the Nb thin film is in a dirty limit \cite{bose2006upper,unpublishedl}.  

The superconducting properties of $\omega$-Nb require further investigation using specialized geometries and samples. Additionally, the existence of $\omega$-Nb phase in Nb thin films may also indicate that this phase is present in Nb thin films for superconducting radiofrequency (SRF) cavities for accelerator applications. As the size of the $\omega$-Nb phase is 10-100 nm in this geometry, which again is comparable to the coherence length of Nb ($\approx$40 nm), local deviations in superconducting properties may allow for magnetic flux penetration into Nb thin film SRF cavities and lead to a degradation in the overall quality factor \cite{carlson2021analysis, pack2020vortex}.

We report the discovery of omega ($\omega$) phase in Nb thin films deposited by high-power impulsed magnetron sputtering (HiPIMS) for superconducting qubits. Internal stresses in Nb thin film likely caused by stress at the Nb/Si interface and impurities localized at the surface may lead to the $\omega$-phase formation in the Nb thin films. The size of the $\omega$-phase varies from 10 to 100 nm, which is comparable to the coherence length of Nb (40 nm), and the volume fraction of the $\omega$-phase is estimated to be $\approx$1\%. While bulk measurements suggests that the critical temperature (T$_{c}$) of the Nb thin film is similar to that observed for bcc Nb (9.3 K), the superconducting properties of $\omega$-Nb require further investigation, particularly, as a recent study implies that such secondary phases can potentially introduce quantum decoherence through pair-breaking\cite{privatecommunicationl}. We also note that the presence of this $\omega$-phase transition in Nb may matter for the Nb thin film superconducting radiofrequency (SRF) cavities, where they could allow for flux penetration and potentially lead to degradation in the quality factor.

\begin{acknowledgments}
We thank to Nathan Sitaraman for the valuable discussion on the superconducting properties of the omega ($\omega$) phase of Nb. 
This material is based upon work supported by the U.S. Department of Energy, Office of Science, National
Quantum Information Science Research Centers, Superconducting Quantum Materials and Systems Center
(SQMS) under contract number DE-AC02-07CH11359.
We thank the Rigetti chip design and fabrication teams for the development and manufacturing of the qubit
devices used in the reported experimental study, and for Rigetti Computing supporting the development of
these devices. This work made use of the EPIC facility of Northwestern University’s NU\textit{ANCE} Center, which has received support from the SHyNE Resource (NSF ECCS-2025633), the IIN, and Northwestern's MRSEC program (NSF DMR-1720139).
\end{acknowledgments}

\setcounter{figure}{0}

\makeatletter 
\renewcommand{\thefigure}{S\@arabic\c@figure}
\makeatother

\beginsupplement

\begin{figure}

   {\includegraphics[width = 3.3 in]{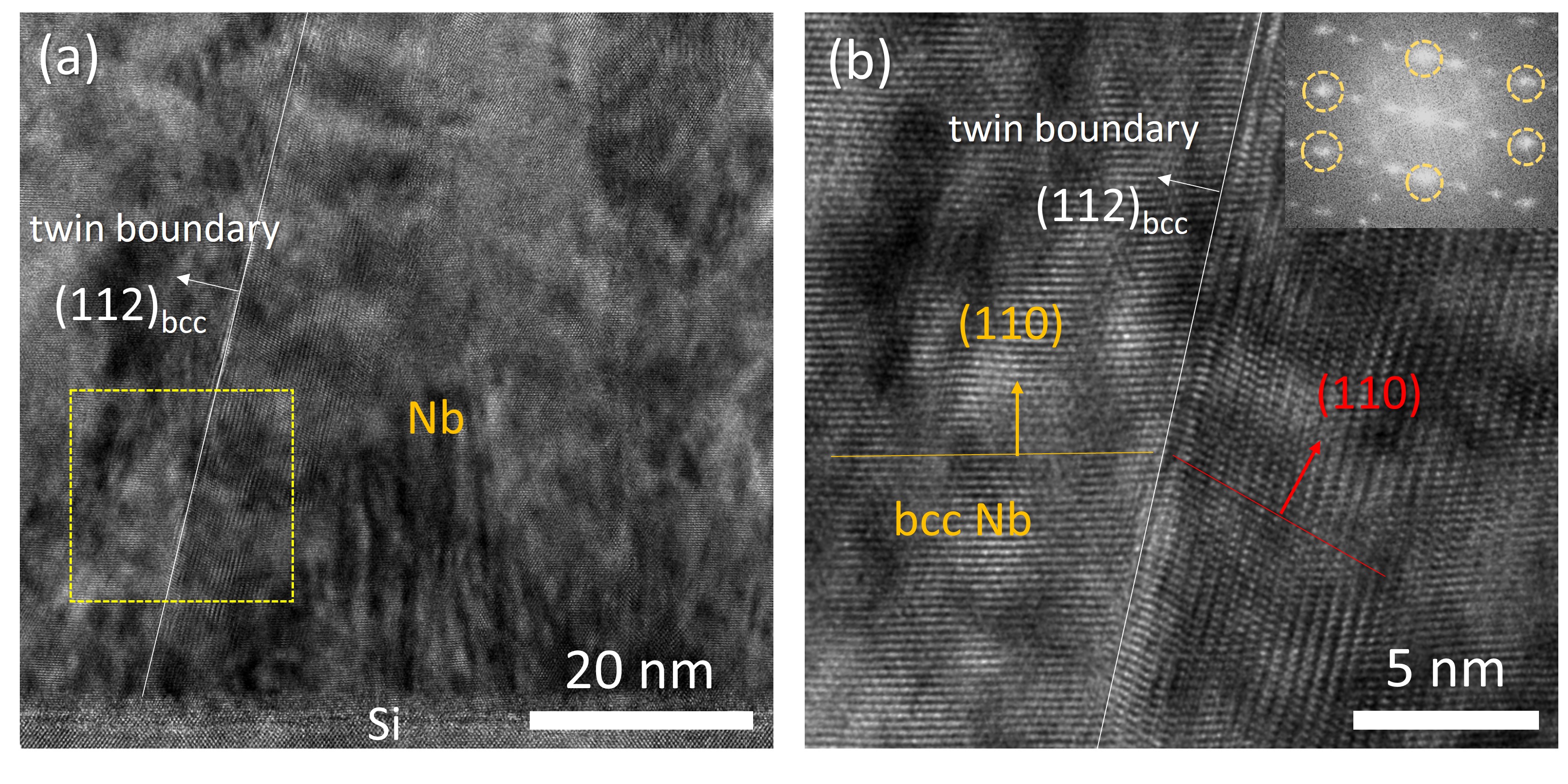}}
   \caption{(a) HR-TEM image of omega phase ($\omega$) in the Nb thin film at Nb/Si interface is provided. (b) The magnified HR-TEM image of $\omega$-Nb phase in Nb at Nb/Si interface from the region denoted by dotted yellow square in Fig. S1(a) is displayed. It shows twin boundary and omega phase transition near the interface, implying that the stresses at the Nb/Si interface may possibly play roles in initiating the $\omega$-phase transition in Nb at the interface.}
\end{figure}

\bibliography{aipsamp.bib}

\end{document}